\documentclass[prl,preprint,twocolumn,superscriptaddress,floatfix,showpacs,10pt]{revtex4-1}%
\usepackage{amsfonts,amsmath,amssymb}
\allowdisplaybreaks
\usepackage[colorlinks=true,urlcolor=blue,anchorcolor=blue,citecolor=blue,filecolor=blue,linkcolor=blue,menucolor=blue,pagecolor=blue,linktocpage=true,pdfproducer=medialab,pdfa=true]{hyperref}
\usepackage{dsfont}
\usepackage{calrsfs}
\DeclareMathAlphabet{\pazocal}{OMS}{zplm}{m}{n}
\usepackage{tikz}
\usepackage{upgreek}

\tikzstyle{dashed}=[dash pattern=on 2pt off 2pt, -, draw=black]
\tikzstyle{small}=[scale=0.6, every node/.style={scale=0.6}]
\tikzstyle{V}=[inner sep=0pt,outer sep=0pt]
\pgfdeclarelayer{nodelayer}
\pgfdeclarelayer{edgelayer}
\pgfsetlayers{nodelayer,edgelayer,main}

\newcommand{\dFsquare}[2]{
\begin{tikzpicture}[baseline={([yshift=-.5ex]current bounding box.center)},small]
  \begin{pgfonlayer}{nodelayer}
    \node[V] (0) at (0.75, 0) {};
    \node (1) at (-.75, 0) {};
    \node[V] (2) at (2, 0) {};
    \node (3) at (-2, 0) {};
  \end{pgfonlayer}
  \begin{pgfonlayer}{edgelayer}
    \draw (3.center) to (1);
    \draw[dashed] (2.center) to (0);
    \ifnum #1 >0 {\draw [dashed] (0) arc (0:-180:0.75);}\else {\draw (0) arc (0:-180:0.75);}\fi
    \ifnum #2 >0 {\draw [dashed] (0) arc (0:180:0.75);} \else {\draw (0) arc (0:180:0.75);}\fi
  \end{pgfonlayer}
\end{tikzpicture}
}

\newcommand{\HCNa}[5] {
\begin{tikzpicture}[baseline={([yshift=-.5ex]current bounding box.center)},small]
  \begin{pgfonlayer}{nodelayer}
    \node (0) at (-2, 0) {};
    \node[V] (1) at (-1.25, 0) {};
    \node[V] (2) at (-0.375, 0) {};
    \node (3) at (2, 0) {};
    \node[V] (4) at (0.375, 0) {};
    \node[V] (5) at (1.25, 0) {};
  \end{pgfonlayer}
  \begin{pgfonlayer}{edgelayer}
    \draw (0.center) to (1);
    \draw[dashed] (5) to (3.center);
    \ifnum #1 >0 {\draw[dashed] (2) arc (0:180:0.875/2);}\else {\draw (2) arc (0:180:0.875/2);}\fi
    \ifnum #2 >0 {\draw[dashed] (1) arc (-180:0:0.875/2);}\else {\draw (1) arc (-180:0:0.875/2);}\fi
    \ifnum #3 >0 {\draw[dashed] (2) to (4);}\else {\draw (2) to (4);}\fi
    \ifnum #4 >0 {\draw[dashed] (5) arc (0:180:0.875/2);}\else {\draw (5) arc (0:180:0.875/2);}\fi
    \ifnum #5 >0 {\draw[dashed] (4) arc (-180:0:0.875/2);}\else {\draw (4) arc (-180:0:0.875/2);}\fi
  \end{pgfonlayer}
\end{tikzpicture}
}
\newcommand{\HCNb}[5] {
\begin{tikzpicture}[baseline={([yshift=-.5ex]current bounding box.center)},small]
	\begin{pgfonlayer}{nodelayer}
		\node (0) at (-2, 0) {};
    \node[V] (1) at (-1, 0) {};
    \node[V] (2) at (0, 1) {};
    \node[V] (3) at (1, 0) {};
		\node (4) at (2, 0) {};
    \node[V] (5) at (0, -1) {};
	\end{pgfonlayer}
	\begin{pgfonlayer}{edgelayer}
		\draw (0.center) to (1);
		\draw[dashed] (3) to (4.center);
    \ifnum #1 >0 {\draw[dashed] (1) arc (0:-90:-1);}\else {\draw (1) arc (0:-90:-1);}\fi
    \ifnum #2 >0 {\draw[dashed] (1) arc (0:90:-1);}\else {\draw (1) arc (0:90:-1);}\fi
    \ifnum #3 >0 {\draw[dashed](2) to (5);}\else {\draw (2) to (5);}\fi
    \ifnum #4 >0 {\draw[dashed] (3) arc (0:90:1);}\else {\draw (3) arc (0:90:1);}\fi
    \ifnum #5 >0 {\draw[dashed] (3) arc (0:-90:1);}\else {\draw (3) arc (0:-90:1);}\fi
  \end{pgfonlayer}
\end{tikzpicture}
}
\newcommand{\HCNc}[5] {
  \begin{tikzpicture}[baseline={([yshift=-1.2ex]current bounding box.center)},small]
	\begin{pgfonlayer}{nodelayer}
		\node (0) at (-2, 0) {};
    \node[V] (1) at (-1, 0) {};
    \node[V] (2) at (-0.5, 1) {};
    \node[V] (3) at (1, 0) {};
		\node (4) at (2, 0) {};
		\node[anchor=north] (5) at (0.5, 1) {};
	\end{pgfonlayer}
	\begin{pgfonlayer}{edgelayer}
		\draw (0.center) to (1);
		\draw[dashed] (3) to (4.center);
    \ifnum #1 >0 {\draw [style=dashed] (1) arc (0:-atan2(1,.56):-1);}\else {\draw (1) arc (0:-atan2(1,.56):-1);}\fi
    \ifnum #2 >0 {\draw [style=dashed] (3) arc (0:-180:1);}\else {\draw (3) arc (0:-180:1);}\fi
    \ifnum #3 >0 {\draw [style=dashed] (5) arc (0:180:.5);}\else {\draw (5) arc (0:180:.5);}\fi
    \ifnum #4 >0 {\draw [style=dashed] (5) arc (0:-180:.5);}\else {\draw (5) arc (0:-180:.5);}\fi
    \ifnum #5 >0 {\draw [style=dashed] (3) arc (0:atan2(1,.56):1);}\else {\draw (3) arc (0:atan2(1,.56):1);}\fi
	\end{pgfonlayer}
\end{tikzpicture}
}

\newcommand{\E}{$\mathcal{E}$}
\newcommand{\scE}{\mathcal{E}}
\newcommand{\scH}{\mathcal{H}}
\newcommand{\scl}{\ell}
\newcommand{\fd}{\mathfrak{d}}
\newcommand{\fo}{\mathfrak{o}}
\newcommand{\DD}{\mathcal{D}}
\newcommand{\fg}{\mathfrak{g}}
\newcommand{\s}{$\sigma$}
\newcommand{\dd}{\mathrm{d}}
\newcommand{\llangle}{\langle\langle}
\newcommand{\rrangle}{\rangle\rangle}
\newcommand{\J}{\mathcal{J}}
\renewcommand{\L}{\mathcal{L}}
\newcommand{\CP}{\mathrm{CP}}
\newcommand{\fgC}{\fg^{\mathds{C}}}
\newcommand{\B}{\upbeta}
\newcommand{\cg}{c_\fg}
\newcommand{\zb}{\overline{z}}
\newcommand{\zetab}{\overline{\zeta}}

\newcommand{\Ah}{\widehat{A}}
\newcommand{\Bh}{\widehat{B}}
\newcommand{\Ch}{\widehat{C}}
\newcommand{\Dh}{\widehat{D}}

\DeclareMathOperator{\res}{res}
\DeclareMathOperator{\Tr}{Tr}

\newcommand{\co}[1]{\parbox{1.7em}{\centering$\displaystyle #1$}}

\begin{document}

\title{RG flow of integrable \E-models}
\preprint{MI-TH-2033}

\author{Falk Hassler}
\email{falk@fhassler.de}
\affiliation{George P. \& Cynthia Woods Mitchell Institute for Fundamental Physics and Astronomy,\\
Texas A\&M University, College Station, TX 77843, USA}

\begin{abstract}
  We compute the one- and two-loop RG flow of integrable \s-models with Poisson-Lie symmetry. They are characterised by a twist function with $2N$ simple poles/zeros and a double pole at infinity. Hence, they capture many of the known integrable deformations in a unified framework, which has a geometric interpretation in terms of surface defects in a 4D Chern-Simons theory. We find that these models are one-loop renormalisable and present a very simple expression for the flow of the twist function. At two loops only models with $N$=1 are renormalisable. Applied to the $\lambda$-deformation on a semisimple group manifold, our results reproduce the $\beta$-functions in the literature.
\end{abstract}

\maketitle

There are two important concepts, integrability and duality, which have led to tremendous insights into the strong coupling region of quantum field theories \cite{Beisert:2010jr}. Both are closely tied to the paramount importance of symmetries in nature. On the one hand, symmetries constrain the dynamics of field theories by imposing that certain observables are conserved. In generic theories the number of independent conserved charges is very small compared to the total count of degrees of freedom. Integrable systems are distinguished by the fact that every degree of freedom comes with an independent conserved charge. Hence, their dynamics is completely governed by symmetry. But this does not render them trivial. A considerable challenge is that the relevant symmetry acts in a highly non-trivial way and identifying it remains a formidable task. Once completed, exact results for all thinkable observables are accessible. Hence, integrable models provide an ideal testing ground for new ideas and concepts in physics. Dualities on the other hand are symmetries on the moduli space of theories. They are versatile and for example relate the strong and the weak coupling regime of particular quantum field theories \cite{Montonen:1977sn}, different space-time geometries in string theory \cite{Buscher:1987sk}, or even gauge theory with quantum gravity \cite{Maldacena:1997re}.

In this letter we are interested in a setup where both, integrability and a duality, called Poisson-Lie (PL) T-duality \cite{Klimcik:1995ux}, go hand in hand. More specifically, we consider a class of 2D, integrable field theories whose couplings describe the geometry of a $D$-dimensional target space. The action describing the corresponding \s-model reads
\begin{equation}\label{eqn:smodel}
  S = \frac1{4\pi\alpha'} \int_\Sigma \left( g_{ij} \dd x^i \wedge \star \dd x^j + B_{ij} \dd x^i  \wedge \dd x^j \right) \,.
\end{equation}
An arbitrary choice of the target space metric $g_{ij}$ and $B$-field $B_{ij}$ does not result in integrability. On the contrary, integrable \s-models are rare and finding them is complicated. Important land marks in the on going search are the principal chiral model (PCM) \cite{Zakharov1978} and the Wess-Zumino-Witten (WZW) model \cite{Novikov:1982ei,*Witten:1983tw,*Witten:1983ar}, both either with a group manifold target space or on a symmetric coset. Deforming them such that integrability is persevered results in two independent families of integrable \s-models, the Yang-Baxter-(or $\eta$-)deformation \cite{Klimcik:2002zj,*Klimcik:2008eq} and the $\lambda$-deformation \cite{Sfetsos:2013wia}. Both admit further integrable deformations on their own and step by step a whole zoo of multi-parameter deformations has been discovered over the last years \cite{Klimcik:2014bta,*Kawaguchi:2014qwa,*Osten:2016dvf,*Borsato:2016pas,*Georgiou:2016urf,*Georgiou:2017oly,*Delduc:2017fib,*Delduc:2018hty,*Georgiou:2018gpe,*Delduc:2019bcl,*Bassi:2019aaf,*Mohammedi:2020qok}. Considering this development one might ask two questions: ``Is there a systematic classification of all the existing examples?'' and ``Is it complete?''. Both are currently subject to active research and therefore not fully answered yet. However, an important observation is that the $\eta$- and $\lambda$-deformation can be related by PL T-duality and an analytic continuation \cite{Hoare:2015gda,*Sfetsos:2015nya,*Klimcik:2015gba}. Similar to integrability, PL T-duality is not applicable to all \s-models \eqref{eqn:smodel} but requires the metric and $B$-field to possess PL symmetry. Remarkably, the $\eta$-, $\lambda$- and many of the multi-parameter deformations have been proven to possess this symmetry. Moreover, they are related by a web of dualities, possibly combined with analytic continuation. Thus, it seems reasonable to expect a deeper connection between integrability and PL T-duality in 2D $\sigma$-models.

Recently, there has been significant progress in formalising this idea. Notably, \cite{Lacroix:2020flf} presents a systematic construction of integrable \E-models \cite{Klimcik:1995dy,Klimcik:1996nq}, which are manifestly PL symmetric, starting from a twist function that governs the infinite tower of conserved currents required by integrability. Their approach is directly related to the construction of integrable models from a 4D Chern-Simons theory \cite{Costello:2017dso,*Costello:2018gyb,*Costello:2019tri}. In this framework, the \E-model describes surface defects and thereby give a geometric interpretation to the twist function whose poles indicate the position of the defects in the normal, 2D space. The objective of this letter is to take a first step from the classical analysis in \cite{Lacroix:2020flf} to the quantum regime. In particular, we calculate the one- and two-loop RG flow of the integrable \E-models they propose based on twist functions with $2N$ simple poles. At one loop these models are renormalisable and the running of the twist function is completely encoded in a meromorphic function $f(z)$, like conjectured by \cite{Delduc:2020vxy} after analysing several examples. At two loops only models with $N=1$ are renormalisable. We interpret this as an possible hint towards quantum corrections that modify the construction of \cite{Lacroix:2020flf} beyond the classical level.

\paragraph*{Integrable \E-models} may be understood as a rewriting of the action \eqref{eqn:smodel} to make PL symmetry manifest. To see how this works, we first change from the Lagrangian perspective to the Hamiltonian \cite{Tseytlin:1990nb,*Tseytlin:1990va}
\begin{equation}\label{eqn:hamilton}
  H = \frac1{4\pi\alpha'} \int \dd\sigma J_M \scH{}^{MN} J_N
\end{equation}
written in terms of the generalised metric
\begin{equation}\label{eqn:genmetric}
  \scH^{MN} = \begin{pmatrix} g_{mn} - B_{mk} g^{kl} B_{ln} \quad &  B_{mk} g^{kn} \\
    -g^{mk} B_{kn}  & g^{mn}  \end{pmatrix}
\end{equation}
and the worldsheet currents $J_M = ( \partial_\sigma x^m \,,\, p_m)$, where $p_m = g_{mn}\partial_\tau x^n + B_{mn} \partial_\sigma x^n$ denotes the canonical momentum of the $\sigma$-model. While the classical, equal time commutation relations that govern the currents
\begin{equation}
  \{ J_M(\sigma), J_N(\sigma') \} = 2 \pi \alpha' \delta'(\sigma-\sigma') \eta_{MN}
\end{equation}
with
\begin{equation}
  \eta_{MN} = \begin{pmatrix} 0 & \delta_n^m \\ \delta_m^n & 0 \end{pmatrix}
\end{equation}
are simple, the Hamiltonian is complicated for generic target space geometries. The appeal of PL symmetry is that it removes this asymmetry. More specifically, it permits a field redefinition,
\begin{equation}
  J_{\Ah} = \frac1{\sqrt{2\pi \alpha'}} E_{\Ah}{}^M(x^i) J_M\,,
\end{equation}
which renders the Hamiltonian quadratic in the currents. A simplification of $H$ is usually penalised by a more complicated current algebra. PL symmetry still guarantees a simple current algebra, namely the classical Ka\v{c}-Moody algebra\cite{Klimcik:2019kkf,*Demulder:2019bha}
\begin{align}
  \{ J_{\Ah}(\sigma), J_{\Bh}(\sigma') \} &=  F_{\Ah\Bh}{}^{\Ch} J_{\Ch} (\sigma) \delta(\sigma-\sigma') + 
    \eta_{\Ah\Bh} \delta'(\sigma-\sigma') \nonumber \\ \label{eqn:emodel}
    \text{with} \quad H &= \frac12 \int J_{\Ah} \mathcal{H}^{\Ah\Bh} J_{\Bh} \,.
\end{align}
Both tensors, $F_{\Ah\Bh}{}^{\Ch}$ and $\eta_{\Ah\Bh}$, on the right hand side of the first equation are independent of target space coordinates $x^i$. Closure of the Poisson bracket requires that $F_{\Ah\Bh}{}^{\Ch}$ are the structure coefficients of a Lie algebra $\fd$ equipped with an ad-invariant, non-degenerate metric $\eta_{\Ah\Bh}$. An index free notation is often convenient. To see how it arises, we introduce the basis $t_{\Ah}$ for $\fd$. Now, the structure coefficients govern the commutator $[t_{\Ah}, t_{\Bh}] = F_{\Ah\Bh}{}^{\Ch} t_{\Ch}$ and the pairing $\llangle t_{\Ah}, t_{\Bh} \rrangle = \eta_{\Ah\Bh}$ implements the metric on $\fd$. The current algebra combined with the Hamiltonian \eqref{eqn:emodel} is called \E-model because the original work \cite{Klimcik:1995dy,Klimcik:1996nq} expresses $H$ in terms of $\scE^{\Ah}{}_{\Bh} = \eta^{\Ah\Ch} \scH_{\Ch\Bh}$ and $\llangle \cdot \,,\, \cdot \rrangle$ instead of using the generalised metric. It is easy to verify from the equations above that $\scE$ is an involution on $\fd$ ($\scE^2 = 1$) and symmetric with respect to the pairing.

PL T-duality arises if different \s-models share the same \E-model \cite{Klimcik:1995ux,Klimcik:1995dy}. For pedagogical reasons, we started from the familiar action \eqref{eqn:smodel} to eventually arrive at \eqref{eqn:emodel}. In practice, going this route is hard. For a given target space geometry it is usually not clear if a generalised frame $E^{\Ah}{}_M$ exists that permits the uplift to an \E-model. Instead one takes the Lie group $\DD$ corresponding to the Lie algebra $\fd$ and constructs for every maximally isotropic subgroup $H_i$ an admissible generalised frame on the coset $H_i\backslash \DD$ \cite{Hassler:2017yza,*Demulder:2018lmj,*Sakatani:2019jgu,*Catal-Ozer:2019hxw,*Hassler:2019wvn,Hassler:2020tvz,*Borsato:2020wwk,*Codina:2020yma}. Remarkably, the scope of this approach is not restricted to PL T-duality but is also essential to find consistent truncations in half-maximal supergravity. Details are not relevant for our discussion. We only need to know that all PL T-dual target space geometries are accessible from the data $\DD$, $H_i$ and $\scE$.

A first hint towards integrable \E-models is that the field equations for the currents $\partial_\tau J^{\Ah} = \{ J^{\Ah}, H \}$ can be written in terms of the flatness condition
\begin{equation}\label{eqn:flatJ}
  \dd \J + \frac12 [ \J, \J ] = 0
\end{equation}
of the $\fd$-valued one-form
\begin{equation}
  \J = \J_\tau \dd\tau + \J_\sigma \dd\sigma = t_{\Ah} \left( \scE^{\Ah}{}_{\Bh} J^{\Bh} \dd\tau + J^{\Ah} \dd\sigma \right)
\end{equation}
on the worldsheet. Equation~\eqref{eqn:flatJ} formally agrees with the flatness condition of a Lax connection $\L(z)$, an essential quantity assigned to every integrable system. However, a closer look reveals two important differences between them. First of all, $\L(z)$ is not a single, but a one parameter family of flat connections. Each member of this family is labeled by the spectral parameter $z$ who takes values in a $\CP^1$. Moreover, while $\J$ is valued in the real Lie algebra $\fd$, the one forms $\L(z)$ return elements of a complexified, semisimple Lie algebra $\fgC$. Intriguingly, it is still possible to relate the two and thereby obtain a large class of integrable \E-models \cite{Severa:2017kcs}. The conserved currents which render the underling PL dual \s-models integrable, can be readily extracted by expanding the Lax connection in the spectral parameter.

As a first step to relate $\J$ with the Lax connection, we have to choose a Lie algebra $\fd$. Following \cite{Lacroix:2020flf}, we identify $\fd = \fg \oplus \dots \oplus \fg$ with $2N$ copies of the Lie algebra $\fg$, that the Lax connection is build on, and label its generators by
\begin{equation}
  t_{\Ah} = t_{\alpha A}\,,
\end{equation}
where $\alpha = 1,\dots,\dim \fg$ and $A = 1, \dots, 2 N$. In the same vein, we define
\begin{equation}\label{eqn:decomp}
  \begin{aligned}
    \scE^{\Ah}{}_{\Bh} &= \delta^\alpha_\beta \scE^A{}_B  &
    \eta_{\Ah\Bh} &= \kappa_{\alpha\beta} \eta_{AB} \\
    F_{\Ah\Bh\Ch} &= f_{\alpha\beta\gamma} F_{ABC}\,.
  \end{aligned}
\end{equation}
Note that we use $\eta_{\Ah\Bh}$ to lower, and its inverse to raise indices. Furthermore, $f_{\alpha\beta\gamma} = f_{\alpha\beta}{}^\delta \kappa_{\delta\gamma}$ denotes the structure coefficients of $\fg$ and $\kappa_{ab}$ its non-degenerate Killing metric. For the latter, we impose the normalisation $f_{\alpha\gamma}{}^\delta f_{\beta\delta}{}^\gamma = - \cg \kappa_{\alpha\beta}$\,. At this point, two important remarks are in order. First, while $\scE^A{}_B$ and $\eta_{AB}$ have exactly the same properties as their counter parts on the full Lie algebra, $F_{ABC}$ is totally symmetric in contrast to the totally antisymmetric $F_{\Ah\Bh\Ch}$ (as a consequence of the ad-invariant pairing). Moreover, we apply a similarity transformation to bring $\scE^A{}_B$ and $\eta_{AB}$ into the canonical form
\begin{equation}\label{eqn:gaugefix}
  \scE^A{}_B = \begin{pmatrix}
    \delta^a_b & 0 \\
    0 & - \delta^{\bar a}_{\bar b}
  \end{pmatrix}
  \quad \text{and} \quad
  \eta_{AB} = \begin{pmatrix}
    \delta_{ab} & 0 \\
    0 & - \delta_{\bar a\bar b}
  \end{pmatrix}
\end{equation}
with bared and unbared lowercase indices running from one to $N$ each. Doing so has the advantage that all relevant data about the \E-model is packaged into a single object, the generalised fluxes $F_{ABC}$. 

We are interested in \E-models whose generalised fluxed are related to the twist function
\begin{equation}
  \varphi(z) = - l_\infty \frac{\prod_{A=1}^{2N} (z - \zeta_A) }{\prod_{A=1}^{2N}(z - z_A)}
\end{equation}
with $2 N$ simple zeros $\zeta_A$, $2 N$ simple poles $z_A$ and one double pole at infinity. The twist function is a powerful tool to capture all relevant information for a large class of integrable \s-models. Its inverse fixes the Maillet $r$-$s$ algebra \cite{Maillet:1985ec,*Maillet:1985ek} of the conserved currents extracted from the Lax connection. As a consequence, all conserved currents are in involution (=independent) and the \s-model is strong integrable. Note that the $r$-$s$ algebra in general does not require a twist function. Hence, we are dealing here still with a special case. Moreover, the twist function might have higher order poles. Corresponding \E-models exist \cite{Lacroix:2020flf}, but for simplicity we restrict the discussion to simple poles. To reproduce the correct signature of $\eta_{AB}$ and $\scE_{AB}$, the additional constraints
\begin{equation}
  \varphi'(\zeta_a) < 0 \quad \text{and} \quad \varphi'(\zeta_{\bar a}) > 0
\end{equation}
have to be imposed. Following the presentation in \cite{Lacroix:2020flf}, one eventually obtains the generalised fluxes
\begin{equation}\label{eqn:FABC}
  F_{ABC} = \sum\limits^{2 N}_{D=1} \frac{L_A L_B L_C K_D}{(z_D - \zeta_A)(z_D - \zeta_B)(z_D - \zeta_C)}\,,
\end{equation}
by diagonalising the $\scE^{\Ah}{}_{\Bh}$ and the pairing $\eta_{\Ah\Bh}$ they present. For the sake of brevity, it is convenient to define $L_A = \left| \varphi'(\zeta_A) \right|^{-1/2}$ and $K_A$, which denotes the residue $\res_{z_A} \varphi$ of the twist function at the pole $z_A$. In combination with \eqref{eqn:decomp} and \eqref{eqn:gaugefix}, they give rise to an integrable \E-model with Lax connection
\begin{equation}
  \L(z) = \sum\limits^{2 N}_{A=1} \frac{L_A \J^A}{z-\zeta_A}\,.
\end{equation}
The one-forms $\J^A = t_\alpha \J^{\alpha A}$ in this expression coincide with the ones in the field equations \eqref{eqn:flatJ}. 

Our discussion so far was only concerned with the classical regime. But clearly the ultimate objective is to quantize the structures we encountered. Unfortunately, it turns out that this task is more complicated than one might initially think. Quantisation of integrable systems usually employs the Quantum Inverse Scattering Method (QISM) \cite{Takhtajan:1979iv,*Faddeev:1979gh}. But the \s-models we encounter here are non-ultra-local, because the twist function's contribution to the $s$ part of the $r$-$s$ algebra does not vanish. Thus, the QISM is not directly applicable to explore the quantum regime. An promising approach to circumvent this problem is an alleviation procedure \cite{Delduc:2012qb,*Delduc:2012vq}, originally inspired by \cite{Faddeev:1985qu}. It has been applied to the $\lambda$-deformation to construct an ultra-local model which can be quantised on the lattice, preserving integrability \cite{Appadu:2017fff,*Appadu:2018ioy}. In the IR limit, the lattice model is argued to be in the same universality class as the original $\lambda$-deformation. This observation emphasis how important it is to understand another quantum effect for these models, their RG flow. Fortunately, it is possible to exploit PL symmetry to extract it up to two loops with minimal effort.

\paragraph*{The one- and two-loop RG flow} of the \s-model \eqref{eqn:smodel} preserves PL symmetry and the \E-model is therefore renormalisable \cite{Sfetsos:2009vt,*Berman:2007xn,*Avramis:2009xi,*Severa:2016lwc,*Pulmann:2020omk,Hassler:2020wnp}. The literature discusses two slightly different approaches to quantify its $\B$-function. One might consider the running of the generalised metric $\mathcal{H}_{\Ah\Bh}$, or equally $\scE^{\Ah}{}_{\Bh}$ since $\eta_{\Ah\Bh}$ does not flow. Alternatively, the generalised metric may be gauged fixed to a standard, diagonal form and instead the generalised frame $E_{\Ah}{}^I$ flows, namely \footnote{$\dot{\scE}^{\Ah}{}_{\Bh} = \beta^{\Ah}{}_{\Ch} \scE^{\Ch}{}_{\Bh} - \scE^{\Ah}{}_{\Ch} \beta^{\Ch}{}_{\Bh}$ yields the flow of \E.}
\begin{equation}
  \frac{\dd E_{\Ah}{}^I}{\dd \log\mu} E_{\Bh I} := \dot{E}_{\Ah}{}^I E_{\Bh I} = \B_{\Ah\Bh} \,.
\end{equation}
The resulting $\B$-function is valued in the Lie algebra $\fo$($D$,$D$) but not all of the $D (2D-1)$ couplings it describes are physically relevant. Due to the global double Lorentz symmetry of the \E-model $D(D-1)$ of them are pure gauge and only the remaining $D^2$ generators of the coset O($D$,$D$)/(O($D$)$\times$O($D$)) have to be considered. Nevertheless, the moduli space of integrable \E-models is in general much smaller and thus furnishes a submanifold $M$ embedded in this coset. We describe the embedding in terms of the Maurer-Cartan form $\Theta: T M \rightarrow \fo(D,D)$, defined by
\begin{equation}
  \dd F_{\Ah\Bh\Ch} = 3 \Theta_{[\Ah}{}^{\Dh} F_{\Bh\Ch]\Dh}\,.
\end{equation}
From the decomposition \eqref{eqn:decomp}, we find $\Theta^{\Ah}{}_{\Bh} = \delta^\alpha_\beta \Theta^A{}_B$, which together with the generalised fluxes in \eqref{eqn:FABC} eventually yields
\begin{equation}
  \Theta_{AB} = \begin{cases}
    \displaystyle \sum\limits_{C=1}^{2 N} \frac{L_A L_B K_C \dd z_C}{(\zeta_A - \zeta_B)(z_C - \zeta_A)(z_C - \zeta_B)} & A \ne B \\
    0 & A = B\,.
  \end{cases}
\end{equation}
Note that $M$ is here parameterised by the $4N+1$ independent coordinates $z_A$, $K_A$ and $l_\infty$. We favour $K_A$ over $\zeta_A$ because they are RG invariants as arbitrary O($D$,$D$) transformations do not affect them. Due to the same mechanism $l_\infty$ does not change under the flow. We normalise it to one by an appropriate rescaling of the spectral parameter $z$. Furthermore, one observes that the center of mass coordinate $Z = \sum_{A=1}^{2N} z_A / (2N)$ does not contribute to $\Theta_{AB}$. Hence, it is one more RG invariant and can be used without loss of generality to set $z_{2N}$=$0$.

The necessary and sufficient condition for our integrable \E-models to be renormalisable is
\begin{equation}\label{eqn:renormal}
  \iota_{\beta} \Theta_{a\bar b} = \B_{a\bar b} \quad \text{with} \quad
  \beta = \sum\limits_{A=1}^{2N} \beta_A \partial_{z_A} \quad \text{and} \quad 
  \dot z_A = \beta_A \,,
\end{equation}
where we take into account the canonical decomposition $\B_{\Ah\Bh} = \kappa_{\alpha\beta} \B_{AB}$. Thanks to the gauge fixing \eqref{eqn:gaugefix}, it is simple to extract the physically relevant $N^2$ coset generators mentioned above: One just has to restrict $A$ and $B$ to $a$ and $\bar b$, respectively. Still beyond $N$=$1$, \eqref{eqn:renormal} is a non-trivial constraint because there are only $2N - 1$ independent variables, $z_A$ with $A$=$1,\dots,2N-1$, but $N^2$ equations. To gain more insights, we have to evaluate the right hand side.

We use the $\alpha'$-expansion
\begin{equation}
  \B_{a\bar b} = \alpha' \B^{(1)}_{a\bar b} + \alpha'^2 \B^{(2)}_{a\bar b} + \mathcal{O}(\alpha'^3)
\end{equation}
to organise the computation. $n$-loop Feynman diagrams with two external legs are a convenient way to present the contributions to $\B^{(n)}_{a\bar b}$. Following \cite{Hassler:2020wnp}, we define the following vertices and propagators:
\begin{equation}
  \begin{aligned}
    F_{abc} &= \tikz[baseline=(b.base)] {
      \draw (0:0) node {} -- (0:0.4) node[anchor=west] (b) {$b$};
      \draw (0:0) -- (120:0.4) node[anchor=east] {$a$};
      \draw (0:0) -- (240:0.4) node[anchor=east] {$c$};
    } &
    F_{\bar abc} &= \tikz[baseline=(b.base)] {
      \draw (0:0) node {} -- (0:0.4) node[anchor=west] (b) {$b$};
      \draw[dashed] (0:0) -- (120:0.4) node[anchor=east] {$\bar a$};
      \draw (0:0) -- (240:0.4) node[anchor=east] {$c$};
    } & 
    \delta^{ab} &= \tikz[baseline=(a.base)]{
      \draw (0,0) node[anchor=east] (a) {$a$} -- (0.5,0) node[anchor=west] {$b$};
    } \\
    F_{a\bar b\bar c} &= \tikz[baseline=(b.base)] {
      \draw[dashed] (0:0) node {} -- (0:0.4) node[anchor=west] (b) {$\bar b$};
      \draw (0:0) -- (120:0.4) node[anchor=east] {$a$};
      \draw[dashed] (0:0) -- (240:0.4) node[anchor=east] {$\bar c$};
    } &
    F_{\bar a\bar b\bar c} &= \tikz[baseline=(b.base)] {
      \draw[dashed] (0:0) node {} -- (0:0.4) node[anchor=west] (b) {$\bar b$};
      \draw[dashed] (0:0) -- (120:0.4) node[anchor=east] {$\bar a$};
      \draw[dashed] (0:0) -- (240:0.4) node[anchor=east] {$\bar c$};
    } &
    \delta^{\bar a\bar b} &= \tikz[baseline=(a.base)]{
      \draw[dashed] (0,0) node[anchor=east] (a) {$\bar a$} -- (0.5,0) node[anchor=west] {$\bar b$\,.};
    }
  \end{aligned}
\end{equation}
Dummy indices and also external indices are suppressed if there are no ambiguities. In this convention, the one-loop $\B$-function incorporates a single diagram,
\begin{equation}\label{eqn:beta1}
  \cg^{-1} \B^{(1)}_{a\bar b} = - \dFsquare01 = - \frac{L_a L_{\bar b}}{(\zeta_a - \zeta_{\bar b})^2}\,.
\end{equation}
Intriguingly, it is possible to find a vector $\beta^{(1)}_A$ that solves \eqref{eqn:renormal} for arbitrary $N$. It can be written in terms of a function $f(z)$ with $N$ simple poles and a double pole at infinity,
\begin{equation}\label{eqn:f(z)}
  f(z) = \sum\limits_{a=1}^N \frac{\res_{\zeta_a} \varphi^{-1}(z)}{z - \zeta_a} - \scl_\infty\,,
\end{equation}
where $\scl_\infty$ is fixed by the constraint $\sum_{A=1}^{2N} f(z_A) = 0$. With $f(z)$ at hand, the one-loop $\beta$-functions of the coordinates $z_A$ have the simple form
\begin{equation}\label{eqn:betaf}
  \beta^{(1)}_A = \cg f(z_A)\,,
\end{equation}
which was already conjectured in \cite{Delduc:2020vxy}. Taking furthermore into account that $K_A$ and $l_\infty$ are invariant, the complete one-loop RG flow of the twist function is summarised by the compact equation
\begin{equation}
  \dot{\varphi}(z) = \cg \sum\limits_{A=1}^{2 N} \frac{f(z_A) K_A}{(z-z_A)^2} + \mathcal{O}(\alpha'^2)\,.
\end{equation}

It is instructive to understand this result in terms of a hierarchy of consistent truncations. In a general $\sigma$-model an infinite number of couplings, the values of the metric and $B$-field at all points in the target space, flow. PL symmetry completely decouples $D^2$ of them and it is save (or more formally consistent) to remove (truncate) the reset of them. This is not the end though. Remember that $D=N \dim \fg$, while we found $2N-1$ running couplings. But $D^2$ is clearly larger for any semisimple Lie algebra $\fg$. Fortunately, the \E-models we consider have an additional symmetry, which is generated by the diagonal subgroup of the Lie group $\DD=G\times \dots \times G$ corresponding to $\fd$ \cite{Lacroix:2020flf}. As a consequence, $\B_{\Ah\Bh}$ has to be invariant under the action of this subgroup and is therefore constrained to $\B_{\Ah\Bh} = \kappa_{\alpha\beta} \B_{AB}$, leaving $N^2$ couplings that might flow. For people working on consistent truncations in supergravity this mechanism is well known and it is not surprising that it applies here, too. However, we have to perform another truncation from $N^2$ to $2N-1$ assuming that $N>1$. It is not possible to identify another, linearly realised symmetry of our \E-models that might facilitate this truncation. Nevertheless, we miraculously succeeded in constructing a further consistent truncation rendering the discussed \E-models one-loop renormalisable.

We now show that this ``miracle'' does not continue to happen for the two-loop RG flow. To this end, we evaluate the $\B$-function \cite{Hassler:2020wnp}
\begin{widetext}
  \begin{equation}
  \begin{aligned}
    2 \cg^{-2} \B^{(2)}_{a\bar b} = 
      &\HCNa00001 \co{+}  \HCNa00101 \co{+} \HCNb00011 \co{+} \HCNb00100 \co{+2} \HCNb01001 \\
      \co{-2} &\HCNb01010 \co{+} \HCNb11000 \co{-2} \HCNc00010 \co{-4} \HCNc00011 \co{+} \HCNc00110 \\
      \co{+4} &\HCNc01010 \co{+} \HCNc10001 \co{-2} \HCNc10010 \co{+} 
      \tikz[baseline={([yshift=-0.5ex]arrow.center)}]{
        \draw (0,0) -- +(0.5,0) node[anchor=west] (arrow) {$\leftrightarrow$};
        \draw[dashed] (arrow.east) -- +(0.5,0);
      } \,.
  \end{aligned}
  \end{equation}
\end{widetext}
It consists of 26 different terms and is even under a $\mathbb{Z}_2$ symmetry flipping dashed and undashed propagators. Hence, we only need to write down half of the terms because all others are fixed by symmetry. For $N$=1, there is only one entry in the $\B$-function, $\B^{(2)}_{1\bar 1}$, and one can easily extract
\begin{equation}\label{eqn:beta2}
  \beta^{(2)} = \cg \beta^{(1)} \frac{(\zeta - \zetab)^2 - 3 (z + \zb)(\zeta + \zetab) + 6 (z\zb + \zeta\zetab)}{2 l_\infty ( \zetab - \zeta )^3}
\end{equation}
with $z_A = ( z\,,\, \zb )$, $\zeta_A = ( \zeta\,,\, \zetab )$ and $\beta_A = ( \beta\,,\, - \beta )$.

For $N>$1, $N^2>2N-1$ and consequentially, the coupling space admits directions normal to the $z_A$'s. For simplicity, we focus here on the simplest case, $N$=2, with one normal direction $y$. In analogy with $\Theta_{a\bar b}$, we denote the corresponding Maurer-Cartan form as
\begin{equation}\label{eqn:Thetaperp}
  \Theta^\perp_{a\bar b} = (-1)^{a+\bar b} \frac{(\zeta_a - \zeta_{\bar b})^2}{L_a L_{\bar b}} \dd y \,.
\end{equation}
It has the defining property
\begin{equation}
  \Tr( \Theta^T \Theta^\perp )= 0\,,
\end{equation}
where $\Theta$ and $\Theta^\perp$ with suppressed indices are treated as 2$\times$2 matrices. One-loop renormalisability for $N$=2 is equivalent to
\begin{equation}
  \Tr ( \B^{(1) T} \Theta^\perp ) = 0
\end{equation}
and can be easily verified by using \eqref{eqn:beta1} and \eqref{eqn:Thetaperp}. For a generic twist function with
\begin{equation}
  z_A = ( 3\,,\,2\,,\,1\,,\,0 )\,, \quad
  \zeta_A = ( -1\,,\,5\,,\,4\,,\,-2)\,, \quad l_\infty =  1 \,,
\end{equation}
one finds
\begin{equation}
  \Tr(\B^{(2) T} \Theta^\perp ) = - \cg^2 \frac{111 2544}{1500625} \dd y \ne 0
\end{equation}
and therefore proves that models with $N>1$ are not two-loop renormalisable. We performed a thorough numerical analysis with different twist functions for $N$=4,6,8 and did not encounter a single configuration with vanishing $\beta$-functions in the normal directions.

There are various ways to interpret this result, among them are:
\begin{itemize}
  \item Integrability is broken by two-loop effects.
  \item One can construct a larger class of integrable \E-models that are incorporated by the RG flow beyond one loop.
  \item The \E-operator, and with it the Lax connection, receive quantum corrections. Saying that they are modified at the order $\alpha'$.
\end{itemize}
It is beyond the scope of this short letter to explore these ideas further. At the moment, the last option seems most plausible. But there is definitely a lot more to learn from the intriguing connection of PL symmetry and integrability. Approaching this question from the 4D Chern-Simons theory description likely provides additional clues.

\paragraph*{The $\lambda$-deformation} on a semisimple Lie group $G$ provides a good crosscheck of our results as its one- and two-loop RG flow has been intensively studied in the literature recently. Its twist function is characterised by
\begin{equation}
  z_A = \frac{2 k \kappa^2}{1- \kappa^2}( -1\,,\,1)\,,\quad
  \zeta_A = \frac{2 k \kappa}{1-\kappa^2}( 1\,,\,- 1 )\,, \quad
  l_\infty = 1\,.
\end{equation}
We use the deformation parameter $\kappa=(1-\lambda)(1+\lambda)^{-1}$  to make contact with \cite{Hoare:2019mcc,Hassler:2020wnp}, but of course all equations presented here can be easily rewritten in terms of $\lambda$. The model has a $\mathbb{Z}_2$ symmetry which simultaneously flips the signs of $\kappa$ and $k$ \cite{Itsios:2014lca}. At the level of the twist function this symmetry is manifest as a relabeling of the zeros and poles. Evaluating \eqref{eqn:f(z)} one finds the function
\begin{equation}
  f(z) = \frac{z(1-\kappa^2) - 2 k \kappa^3}{2 z (\kappa^2 - 1) + 4 k \kappa}\,,
\end{equation}
controlling the one-loop RG flow. Note that this function differs from the one presented in \cite{Delduc:2020vxy} for the $\lambda$-deformation. It only has one simple pole instead of two. Finally, the RG invariants $K_A = k ( -1\,,\, 1 )$ encode the level of the WZW-model at the origin of the deformation $\kappa$=1. Form \eqref{eqn:betaf}, we extract the one-loop contribution to the $\beta$-function $\beta_A =( \beta\,,\,-\beta )$ for $z_A$,
\begin{equation}
  \beta^{(1)} = \cg f(z) = - \frac{\cg\kappa}2\,.
\end{equation}
At two loops, \eqref{eqn:beta2} yields
\begin{equation}
  \beta^{(2)} = - \cg^2 \frac{(1 - \kappa^2)(3 \kappa^2 + 1)}{32 k}\,.
\end{equation}
To compare our results with the literature, we note that $\beta$ can be alternatively expressed in terms of the $\beta$-function for $\kappa$, 
\begin{equation}
  \beta = \beta_\kappa \frac{\dd z}{\dd \kappa}\,,
\end{equation}
after taking into account that $k$ is not affected by the flow. Inverting this relation, we eventually find the expected \cite{Hoare:2019mcc,Georgiou:2019nbz,Hassler:2020wnp}
\begin{equation}
\beta_\kappa = \cg \frac{(\kappa^2 - 1)^2}{8 k} \left( 1 +  \cg \frac{1+2\kappa^2-3\kappa^4}{16 k\kappa} \right)\,.
\end{equation}

\begin{acknowledgments}
\paragraph*{Acknowledgements:}
We thank D. Butter, G. Piccinini, C. Pope and D. Thompson for inspiring discussions. Furthermore, we acknowledge the online seminar series ``Integrability, Dualities and Deformations'' for its continuous supply of interesting talks and discussions about PL symmetry and integrable deformations.
\end{acknowledgments}

\bibliographystyle{apsrev4-1}
\bibliography{literature}

\end{document}